\documentclass[12pt]{iopart}
\expandafter\let\csname equation*\endcsname\relax
\expandafter\let\csname endequation*\endcsname\relax
\usepackage{graphicx}
\usepackage{amsmath}

\usepackage{color}

\begin{document}

\title{Theory for the single-particle dynamics in 
glassy mixtures with particle size swaps}

\author{Grzegorz Szamel}
%\affiliation{
\address{
Department of Chemistry,
Colorado State University, Fort Collins, CO 80523}

\date{\today}

\begin{abstract}
We present a theory for the single-particle dynamics in binary mixtures 
with particle size swaps. The general structure of the theory follows that 
of the theory for the collective dynamics in binary mixtures with particle size swaps,
which we developed previously [G. Szamel, Phys. Rev. E \textbf{98}, 050601(R) (2018)].
Particle size swaps open up an additional relaxation channel, which
speeds up both the collective dynamics and the single-particle dynamics. 
To make explicit predictions, we resort to a factorization approximation 
similar to that employed in the mode-coupling theory of glassy dynamics. 
We show that, like in the standard mode-coupling theory, the single-particle 
motion becomes arrested at the dynamic glass transition predicted by
the theory for the collective dynamics. We compare the non-ergodicity 
parameters predicted by our mode-coupling-like approach for the an equimolar
binary hard sphere mixture with particle size swaps with the non-ergodicity 
parameters predicted by the standard mode-coupling theory for the same system 
without swaps. Our theory predicts that the ``cage size'' is bigger in the
system with particle size swaps.
\end{abstract}

\maketitle

\section{Introduction}

Computer simulations are an essential tool in the quest for the understanding of 
glassy dynamics and the glass transition \cite{BerthierBiroliRMP}. 
Their usefulness comes from the fact that they can access particles' 
positions and thus are able to provide much
more detailed picture of the dynamics than even most ingenious experiments. 
However, for many years computer simulations had to cope with the ``glass ceiling''
\cite{BerthierPNAS}: since timescales accessible in simulations are orders
of magnitude shorter than experimental timescales, computer simulations utilizing
well equilibrated systems were only feasible in the temperature/density region
where the dynamics are only 4-5 decades slower than in normal stable liquids. 

Recently, a creative extension of the particle size swaps Monte Carlo,
that was first used in the context of glassy systems' simulations almost 30 years
ago \cite{Grigera}, was introduced. We should recall that since the original study
of Ref. \cite{Grigera}, the particle size swaps Monte Carlo method has been tried 
a few times \cite{Flenner2006}, but with very limited success. Then, Berthier and 
collaborators \cite{BerthierCNO,NinarelloPRX} turned the problem on its head 
by introducing several new glassy 
systems for which the particle size swaps Monte Carlo method speeds up the 
equilibration by many orders of magnitude. It is now possible to equilibrate 
glassy systems at temperatures comparable to and even lower than the laboratory glass
transition temperature. This allowed for the investigation of 
static properties of glassy systems at realistic conditions 
\cite{BerthierPNAS,OzawaPNAS,WangNatComm,Berthier2d}.

The success of the particle size swaps method opened two sets of theoretical questions: 
questions about the method itself and questions about our fundamental understanding 
of the glass transition. The latter questions are concerned with the explanation
of the success of the particle size swaps method. While it is intuitively plausible
that particle size swaps open a new relaxation channel and thus should speeding up
the dynamics, the amount of the speed-up is quite significant and calls for 
a more detailed explanation. Additionally, one would like to understand theoretically
when does the particle size swaps method work and when it does not. Finally, 
one would like to describe the eventual slowing down of the particle size swaps
Monte Carlo method and compare it with the slowing down of local dynamics, either
Monte Carlo, or Brownian, or Newtonian dynamics.   

Out of the second set of questions opened by the success of the particle size 
swaps Monte Carlo, probably the most crucial one is concerned with the importance  
of cooperative processes operating on length scales that increase upon approaching 
the glass transition compared to the importance of local dynamics' slowing down 
\cite{WyartCates,BBBT}. Additional questions are concerned with the relevance 
of configurational entropies calculated theoretically by introducing order parameters
that explicitly forbid or allow particle size swaps 
\cite{IkedaZamponiIkeda,IkedaZamponi}.

In spite of these interesting questions raised by the success of the particle size 
swaps Monte Carlo method, the theoretical analysis of this method is in its infancy. 
Three approaches have been proposed. First, Ikeda \textit{et al.} 
\cite{IkedaZamponiIkeda,IkedaZamponi} proposed to use the close relationship between
the dynamics of mean-field models and the replica theory-based description of
these models. Specifically, they used the correspondence between the dynamic 
transition predicted by the dynamic (mode-coupling-like) theory and identified with the 
dynamic crossover observed in simulations and the dynamic transition predicted by 
the replica approach. They argued that in the presence of particles size swaps
one has to adopt a more general structure of the \textit{ansatz} for the 
inter-replica correlation functions, which are the order parameters within the
replica approach. They showed that the new \textit{ansatz} leads to a shift of the
dynamic transition towards larger volume fractions, in qualitative agreement
with the simulations.

Next, Brito \textit{et al.} \cite{BritoLernerWyart}, in the context of continuous
distribution of particles' diameters, proposed to treat the diameters as additional 
variables, endowed with their own equations of motion (stochastic in the case of 
the Monte Carlo dynamics and deterministic continuous time equations in the case of
the Newtonian-like dynamics). This allowed Brito \textit{et al.} to put forward some 
general arguments, which imply that allowing particle size swaps should lead to the 
decrease of the onset temperature for the glassy dynamics. However, their arguments 
do not lead to any quantitative prediction for this change nor to any  
prediction for the actual dynamical behavior with particle size swaps. Nevertheless,
Brito \textit{et al.}'s idea of treating particles' diameters as additional variables
is extremely useful and it has already lead to generalizations of the original
particle size swaps Monte Carlo method \cite{Kapteijns,MDswap}. 

We developed a \textit{dynamic} theory for the acceleration due to the particle size
swaps \cite{Szamelswap}. 
The general structure of our theory agrees with the physical intuition:
particle size swaps act as an additional relaxation channel, which speeds up
the dynamics. To make explicit predictions we had to approximate the so-called 
irreducible memory functions, which are the main unknown theoretical quantities
in our approach. To this end we used a factorization approximation of the type
utilized in the mode-coupling theory of glassy dynamics and the glass
transition \cite{Goetzebook}. We calculated an approximate glass transition
phase diagram  for an equimolar binary hard-sphere mixture. The presence of particle 
size swaps shifts the dynamic glass transition towards higher volume fractions.
The shift increases with increasing ratio of the hard-sphere diameters; 
it saturates at about 4\% at the diameter ratio of approximately 1.2.
We would like to emphasize that in our theory the set of order parameters
does not depend on whether particle size swaps are allowed or not. It is the
set of self-consistent equations for the order parameters that changes. This
is in contrast with the approach of Ikeda \textit{et al.} 

Here we extend the previously presented theory to the case of single-particle
motion. This extension will allow us to study the frozen-in part of the self-van 
Hove function
(albeit in the reciprocal space) and compare the localization lengths with
and without particle size swaps. Notably, the set of order parameters for the 
single-particle motion does depend on whether particle size swaps are allowed or not.
This is due to the fact that without particle size swaps there cannot be any
single-particle cross-correlations. 

Generally speaking, the relation of the structure of the theory for the single-particle
motion and the previously derived theory for the collective dynamics is very similar
to the relation between the same theories within the standard mode-coupling 
theory of the glass transition. In particular, in the present case involving
particle size swaps and in the standard theory describing local dynamics,
single-particle motion becomes arrested at the dynamic glass transition 
predicted by the theory for collective dynamics.

The paper is organized as follows. In the next section we discuss the model
system with particle size swaps that we analyze and introduce the self-intermediate
scattering functions that quantify the single-particle dynamics. In Sec. 3 we 
express these functions in terms of the so-called memory functions. In Sec. 4
we introduce the main approximation of the theory, the factorization approximation,
and use it to express the memory functions in terms of the intermediate
scattering functions. In Sec. 5 we present numerical results for the non-decaying
parts of the intermediate scattering functions in the glassy phase predicted by the 
theory. We end with some discussion in Sec. 6. 

\section{Model: Binary mixture with particle size swaps}

We consider the model investigated in Ref. \cite{Szamelswap}, \textit{i.e.} a binary 
mixture. This is the simplest model that allows one to investigate the influence 
of the particle size swaps on the dynamics. Two arguments can be made against
using a binary mixture. First, computer simulations studies \cite{NinarelloPRX} showed 
that particle size exchanges in systems with a continuous polydispersity result
in the largest speed-up of the dynamics. Second, a system with a continuous 
polydispersity allows one to treat the particles' diameters as additional variables 
that evolve according to their own equation of motion \cite{BritoLernerWyart}.
The distribution of the particles diameters is then enforced by a diameter-dependent 
chemical potential-like term, which plays the role of an one-particle external 
potential for the diameter variable. The evolution of particles positions and 
sizes could then be implemented in a very similar way. However, the price that one has 
to pay for this uniformity of the evolution equations is that one has to deal with
a static problem with infinitely many components. This makes explicit calculations
a bit daunting.  

To investigate the single-particle dynamics we consider a binary mixture consisting 
of $N$ particles in volume $V$. We assume that one of these particles, for convenience
the particle number 1, is tagged and focus our attention on the motion of this particle.
All the particles, including the tagged particle, can be of type $A$ or $B$.
These particles types differ by size. Since any particle 
can change its type, the state of particle $i$ is determined by its position,
$\mathbf{r}_i$, and type indicated by a binary variable $\sigma_i$,
with $\sigma_i=1$ corresponding to $A$ and $\sigma_i=-1$ corresponding to $B$. 
The composition of the system is specified by the difference of the chemical
potentials of particles of type $A$ and $B$, $\Delta \mu$. We note that for 
a given chemical potential difference, the
composition depends also on the number density $n=N/V$ and the temperature $T$
of the system. We will assume that these three parameters result in 
concentrations $x_A$ and $x_B=1-x_A$, and that these concentration are constant while
the density or the temperature of the system varies. We note that this implies that
the tagged particle can be of type $A$ with probability $x_A$ and of type $B$ with
probability $x_B$. In practical calculations 
we will restrict ourselves to equimolar mixtures (for which some formulas simplify).
We assume that the ``normal'' dynamics of the system is Brownian, \textit{i.e.}, that 
each particle moves under the combined influence of thermal noise and interparticle 
forces. The forces are derived from a spherically symmetric potential, which depends 
on the particle type, $V_{\sigma_i \sigma_j}(r_{ij})$, where 
$r_{ij}=|\mathbf{r}_i-\mathbf{r}_j|$ is the distance between particles $i$ and $j$.
In addition to the Brownian motion in space, the particles can change their type (size). 
We assume that each particle can change its type independently of the type changes
of other particles. This is analogous to the single-spin-flip dynamics of spin
systems. In contrast, in the initial computational 
studies \cite{BerthierCNO,NinarelloPRX} particle size swaps were done 
in such a way that the number of particles of each
size was conserved. This procedure is analogous to the so-called Kawasaki dynamics 
of spin systems. It is convenient in simulational studies because it allows one to 
maintain easily a specific composition of the system. Later investigations
\cite{Kapteijns,MDswap} showed that replacing the Kawasaki-type procedure by 
single particle size swaps is a relatively mild change 
and that both procedures lead to very similar results. 

The above described model corresponds to the following equation of motion 
for the $N$-particle distribution,
$P_N(\mathbf{r}_1,\sigma_1, ..., \mathbf{r}_N,\sigma_N;t)$, abbreviated below as $P_N$,
\begin{eqnarray}\label{Peom}
\partial_t P_N =
\Omega_{\text{sw}}
P_N \equiv \left(\Omega + \delta \Omega_{\text{sw}}\right) P_N.
\end{eqnarray}
Here the evolution operator $\Omega_{\text{sw}}$ consists of two parts
describing two relaxation channels, the part 
describing Brownian motion of particles,
\begin{eqnarray}\label{Omega}
\Omega = D_0 \sum_i \partial_{\mathbf{r}_i}\cdot
\left(\partial_{\mathbf{r}_i}-\beta\mathbf{F}_i\right)
\end{eqnarray}
and the part describing particle size swaps,
\begin{eqnarray}\label{deltaOmegasw}
\delta\Omega_{\text{sw}} = 
- \tau_{\text{sw}}^{-1} \sum_i \left(1-S_i\right) w_i .
\end{eqnarray}
In Eq. (\ref{Omega}) $D_0$ is the diffusion coefficient of an isolated particle,
$D_0=k_BT/\xi_0$, with $\xi_0$ being the friction coefficient of an isolated particle,
$\beta=1/k_B T$, and 
$\mathbf{F}_i$ is the total force on particle $i$, 
$\mathbf{F}_i=\sum_{j\neq i} \mathbf{F}_{ij} = - \sum_{j\neq i} \partial_{\mathbf{r}_i}
V_{\sigma_i\sigma_j}(r_{ij})$. In Eq. (\ref{deltaOmegasw})
$\tau_{\text{sw}}$ is the rate of attempted particle size swaps, $S_i$ is the swap 
operator, $S_i \sigma_i = -\sigma_i$, and $w_i$ is the factor ensuring that the 
detailed balance condition, $\left(1-S_i\right)w_i P_N^{eq} = 0$, is satisfied,
with $P_N^{eq}$ being the equilibrium distribution,
$P_N^{eq}\propto \exp\left(-\beta\sum_{i\neq j} V_{\sigma_i\sigma_j}(r_{ij}) + 
\sum_i \frac{\beta}{2}\Delta\mu\sigma_i \right)$. 
The factor $w_i$ depends on the way particle size swaps are attempted. In practical
applications one typically uses Metropolis criterion for accepting attempted swaps. 
It should be emphasized that while the interactions influencing particles' motion
in space are pairwise-additive, the factor $w_i$ typically is not and it depends on
the whole neighborhood of particle $i$. 

The focus of the present investigation are the elements of the matrix
of the self-intermediate correlation functions, \textit{i.e.} the 
single-particle (tagged particle) density correlation functions,
\begin{eqnarray}\label{Fsabdef}
F_{\alpha\beta}^s(q;t)  
= \left<n_\alpha^s(\mathbf{q})e^{\Omega_{\text{sw}}t} 
n_\beta^s(-\mathbf{q})\right>,
\end{eqnarray}
where $n_{\alpha}^s$ is the Fourier transform of the microscopic 
tagged particle density, 
\begin{equation}\label{nsabdef}
n_{\alpha}^s(\mathbf{q}) 
=   
\frac{1+\sigma_1\left(\delta_{\alpha A}-\delta_{\alpha B}\right)}{2} 
e^{-i\mathbf{q}\cdot\mathbf{r}_1}.
\end{equation}

To develop a theory for the dynamics of $F_{\alpha\beta}^s(q;t)$ we will also
need the collective density correlation functions,
\begin{eqnarray}\label{Fabdef}
F_{\alpha\beta}(q;t)  
= \left<n_\alpha(\mathbf{q})e^{\Omega_{\text{sw}}t} 
n_\beta(-\mathbf{q})\right>,
\end{eqnarray}
where $n_{\alpha}$, $\alpha=A,B$, 
are the Fourier transforms of the normalized microscopic densities
of particles of type $\alpha$,
\begin{equation}\label{nabdef}
n_{\alpha}(\mathbf{q}) 
=  \frac{1}{\sqrt{N}} 
\sum_{i>1} \frac{1+\sigma_i\left(\delta_{\alpha A}-\delta_{\alpha B}\right)}{2} 
e^{-i\mathbf{q}\cdot\mathbf{r}_i}.
\end{equation}
In Eqs. (\ref{Fsabdef},\ref{Fabdef}) 
and in the following equations the standard conventions apply: 
$\left<\dots\right>$ denotes the semi-grand canonical ensemble average
over $P_N^{eq}$, the equilibrium probability distribution
stands to the right of the quantity being averaged, and all operators
act on it as well as on everything else.

We should emphasize the contrast between the single-particle and collective density 
correlation functions. In the absence of particle size swaps the matrix
of the self-intermediate correlation functions is diagonal and the following
notation is used $F_\alpha^s(q;t)\equiv F_{\alpha\alpha}^s(q;t)$, $\alpha=A,B$.
In the presence of particle size swaps the tagged particle can change its type,
which leads to the appearance of the off-diagonal terms in the  matrix
of the self-intermediate correlation functions. In contrast, 
the set of collective density correlation functions is the same regardless of the
presence of particle size swaps. 

In the approach of Ikeda \textit{et al.}, the order parameters are the replica theory
analogues of the single-particle density correlations (physically, the long-time
limits of the time-dependent single-particle density correlation functions should
be equal to the replica theory single-particle correlations; note that a truly
consistent dynamic and static description of the glass transition is still
lacking, except in infinite spatial dimension). Thus, one could argue that it
is natural that in that approach a different set of order parameters is used
depending on the presence of particle size swaps. 

\section{Memory function representation}

To derive the memory function representation of the tagged particle
density correlations we use the standard projection operator procedure
\cite{HansenMcDonald}. First, we define a projection operator on the tagged particle
density subspace, $\mathcal{P}^s$,
\begin{eqnarray}\label{projsdef}
\mathcal{P}^s = \dots \sum_{\alpha} \left. n_\alpha^s(-\mathbf{k})\right>x_\alpha^{-1}
\left< n_\alpha^s(\mathbf{k}) \right. \dots .
\end{eqnarray}
and the orthogonal projection, $\mathcal{Q}$,
\begin{eqnarray}\label{oprojdef}
\mathcal{Q}^s = \mathcal{I} - \mathcal{P}^s \equiv 
\mathcal{I} - \dots \sum_{\alpha} \left. n_\alpha^s(-\mathbf{k})\right>x_\alpha^{-1}
\left< n_\alpha^s(\mathbf{k}) \right. \dots .
\end{eqnarray}
Note that $\left<n_\alpha^s(\mathbf{k}) n_\beta^s(-\mathbf{k})\right> \equiv 
F_{\alpha\beta}^s(\mathbf{k};t=0)=x_\alpha \delta_{\alpha\beta}$.

Next, we use the standard projection operator identity \cite{Goetzebook}
\begin{eqnarray}\label{projident1}
\frac{1}{z-\Omega_\text{sw}} = \frac{1}{z - \Omega_\text{sw}\mathcal{Q}} +
\frac{1}{z - \Omega_\text{sw}\mathcal{Q}}\Omega_\text{sw}\mathcal{P} 
\frac{1}{z-\Omega_\text{sw}}.
\end{eqnarray}
to express the Laplace transform of the time derivative of the tagged particle 
density correlation function,
$zF_{\alpha\beta}^s(q;z)-F_{\alpha\beta}^s(q;t=0)$, 
in terms of the tagged particle reducible memory function,
\begin{eqnarray}\label{projident2}
&& zF_{\alpha\beta}^s(q;z)-F_{\alpha\beta}^s(q;t=0)
= \left<n_\alpha^s(\mathbf{q}) \Omega_\text{sw} 
\frac{1}{z-\Omega_\text{sw}} n_\beta^s(-\mathbf{q}) \right> 
\nonumber \\ &=& 
\left<n_\alpha^s(\mathbf{q}) \Omega_\text{sw} 
\left( \mathcal{P}^s +  \mathcal{Q}^s \right)
\frac{1}{z-\Omega_{sw}} n_\beta^s(-\mathbf{q}) \right>
\nonumber \\ &=& 
\sum_{\gamma} 
\left<n_\alpha^s(\mathbf{q}) \Omega_\text{sw} n_\gamma^s(-\mathbf{q})\right>x_\gamma^{-1}
\left< n_\gamma^s(\mathbf{q}) \frac{1}{z-\Omega_\text{sw}} 
n_\beta^s(-\mathbf{q}) \right> 
\nonumber \\ && + 
\sum_{\gamma} \left<n_\alpha^s(\mathbf{q}) \Omega_\text{sw} \mathcal{Q}^s  
\frac{1}{z - \Omega_\text{sw}\mathcal{Q}^s} 
\mathcal{Q}^s \Omega_\text{sw} n_\gamma^s(-\mathbf{q})\right>x_\gamma^{-1}
\left< n_\gamma^s(\mathbf{q})\frac{1}{z-\Omega_\text{sw}}
n_\beta^s(-\mathbf{q}) \right>.
\nonumber \\
\end{eqnarray}
In the second line of Eq. (\ref{projident2}) we identify the tagged particle 
frequency matrix,
\begin{eqnarray}\label{Omegamsdef}
O_{\alpha\beta}^s(q) &=& 
- \left<n_\alpha^s(\mathbf{q}) \Omega_\text{sw} n_\beta^s(-\mathbf{q})\right> =
D_0 \mathbf{q}^2 x_\alpha \delta_{\alpha\beta} +
\tau_\text{sw}^{-1}\left<n_\alpha^s(\mathbf{q})
\left(1-S_1\right) w_1 n_\beta^s(-\mathbf{q})\right>
\nonumber \\ &=&
D_0 \mathbf{q}^2 x_\alpha \delta_{\alpha\beta} +  
\frac{1}{2\tau_\text{sw}}\left< w_1 \right> \left(2\delta_{\alpha\beta}-1\right),
\end{eqnarray}
and in the third line we identify the matrix of the tagged particle reducible  
memory functions,
\begin{eqnarray}\label{mfreds1a}
M_{\alpha\beta}^{s \text{red}}(q;z) &=& 
\left<n_\alpha^s(\mathbf{q}) \Omega_\text{sw} \mathcal{Q}^s 
\left(z-\mathcal{Q}^s\Omega_{\text{sw}}\mathcal{Q}^s\right)^{-1}
%\frac{1}{z - \mathcal{Q}^s\Omega_\text{sw}\mathcal{Q}^s} 
\mathcal{Q}^s \Omega_\text{sw} n_\beta^s(-\mathbf{q})\right>.
\end{eqnarray}

We note that, like in the collective dynamics problem \cite{Szamelswap}, 
the tagged particle frequency matrix can be decomposed into the part corresponding
to ``normal'' (\textit{i.e.} Brownian) and the part describing particle size swaps.
This separation can be conveniently expressed by introducing 3-dimensional vectors 
$\mathsf{v}_\alpha$, $\alpha=A,B$, and a 3x3 matrix $\mathsf{O}^s$,
\begin{eqnarray}\label{Os33}
O_{\alpha\beta}^s(q) &=& \mathsf{v}_{\alpha}^{\text{T}}
\mathsf{O}^s(q) \mathsf{v}_{\beta}
\end{eqnarray}
Here $\mathsf{v}_{A}^{\text{T}}=(1,0,1)$, $\mathsf{v}_{B}^{\text{T}}=(0,1,-1)$, 
$\mathsf{O}_{11}^s=D_0q^2 x_A$, $\mathsf{O}_{22}^s=D_0q^2 x_B$, 
$\mathsf{O}_{33}^s=(1/2\tau_{\text{sw}})\left<w_1\right>$ and 
$\mathsf{O}_{ab}=0$ for $a\neq b$. Thus, Brownian dynamics is represented by 
elements of matrix $\mathsf{O}^s$ with $a,b \le 2$ and the particle size swaps 
are represented by the 33 element of $\mathsf{O}^s$. 

Next, we note the structure of the ``vertexes'' in the 
matrix of the tagged particle reducible memory functions,
\begin{eqnarray}\label{Ms22vert}
\left. \mathcal{Q}^s \Omega_\text{sw} n_\alpha^s(-\mathbf{q})\right> &=& 
\left. D_0
\mathcal{Q}^s i \mathbf{q}\cdot\left( i\mathbf{q} + \beta\mathbf{F}_1\right)
\frac{1+\sigma_1\left(\delta_{\alpha A}-\delta_{\alpha B}\right)}{2}
e^{i\mathbf{q}\cdot\mathbf{r}_1}\right>
\nonumber \\ &&
- \left. \frac{1}{\tau_{\text{sw}}} \mathcal{Q}^s 
w_1 \sigma_1 e^{i\mathbf{q}\cdot\mathbf{r}_1}
\right>\left(\delta_{\alpha A}-\delta_{\alpha B}\right)
\nonumber \\ &=& 
\sum_a \left. \pi_a^s(-\mathbf{q})\right> \mathsf{v}_{\alpha a},
\end{eqnarray}
where functions $\pi_a^s$ are defined as follows,
\begin{eqnarray}\label{pis12}
\pi_{1,2}^s(\mathbf{q}) \!\! &=& \!\! 
D_0
\mathcal{Q}^s i\mathbf{q}\cdot\left( i\mathbf{q} - \beta\mathbf{F}_1\right)
\frac{1\pm\sigma_1}{2}e^{-i\mathbf{q}\cdot\mathbf{r}_1},
\\ \label{pis3}
\pi_3^s(\mathbf{q}) \!\! &=& \!\! - \frac{1}{\tau_{\text{sw}}} \mathcal{Q}^s 
w_1 \sigma_1 e^{-i\mathbf{q}\cdot\mathbf{r}_1}.
\end{eqnarray}
and $\mathsf{v}_{\alpha a}$ is the $a$th element of vector $\mathsf{v}_{\alpha}$.

The structure of the vertexes uncovered in Eq. (\ref{Ms22vert}) 
allows us to express the matrix of the tagged particle reducible
memory functions $M^{s \text{red}}$ in terms of  3-dimensional vectors 
$\mathsf{v}_\alpha$, $\alpha=A,B$, 
and a 3x3 matrix $\mathsf{M}^{s \text{red}}$,
\begin{eqnarray}\label{Ms33}
M_{\alpha\beta}^{s \text{red}}(q;z) 
&=& \mathsf{v}_{\alpha}^{\text{T}}\mathsf{M}^{s \text{red}}(q;z) 
\mathsf{v}_{\beta},
\end{eqnarray}
where the matrix elements of $\mathsf{M}^{s \text{red}}$ read 
\begin{eqnarray}\label{Ms33red}
\mathsf{M}_{ab}^{s \text{red}}(q;z) = 
\left<\pi_a^s(\mathbf{q}) 
\left(z-\mathcal{Q}^s\Omega_{\text{sw}}\mathcal{Q}^s\right)^{-1}
\pi_b^s(-\mathbf{q})\right>.
\end{eqnarray}

Again, this representation allows us  
to separate ``normal'' (\textit{i.e.} Brownian) dynamics represented by 
elements of matrix $\mathsf{M}^{s \text{red}}$ with $a,b \le 2$ 
and the particle swaps represented by the 33 element
of $\mathsf{M}^{s \text{red}}$. We note, however, that while matrix $\mathsf{O}^s$
is diagonal, in general, matrix $\mathsf{M}^{s \text{red}}$ is not. Thus, 
in general, there will be terms describing the time-delayed couplings between 
the two relaxation channels. 
These couplings are described by elements $\mathsf{M}^{s \text{red}}$ with
$a \le 2$ and $b=3$, and $a=3$ and $b\le 2$. 

Finally, we introduce the tagged particle irreducible evolution operator. This 
will allow us to introduce a 3x3 matrix $\mathsf{M}^{s \text{irr}}$, 
whose elements are functions evolving with the so-called irreducible evolution operator.
For systems evolving with Brownian dynamics this step was introduced and justified 
by Cichocki and Hess \cite{CHess}. Later, Kawasaki \cite{Kawasaki} argued that an
analogous construction should also be applied for a more general class
of systems evolving with stochastic dynamics, including spin systems evolving with 
spin-flip dynamics. We recall that our dynamics is a combination of Brownian
dynamics and particle size swaps, and the latter events are technically implemented 
in terms of single spin flips.

We follow the definition of the collective dynamics irreducible evolution 
operator \cite{Szamelswap} and define the tagged particle irreducible 
evolution operator as follows,
\begin{eqnarray}\label{projSmols1}
\Omega_\text{sw}^{s \text{irr}} &=& \mathcal{Q}^s\Omega_\text{sw}\mathcal{Q}^s -
\sum_{\alpha\beta} \left. \mathcal{Q}^s \Omega n_\alpha^s(-\mathbf{q})\right>
\left<n_\alpha^s(\mathbf{q})\Omega n_\beta^s(-\mathbf{q})\right>^{-1}
\left<n_\beta^s(\mathbf{q})\Omega
\mathcal{Q}^s\right. 
\nonumber \\ && 
-
\left. \mathcal{Q}^s \delta\Omega_\text{sw}\sigma^s(-\mathbf{q})\right>
\left<\sigma^s(\mathbf{q})\delta\Omega_\text{sw}\sigma^s(-\mathbf{q})\right>^{-1}
\left<\sigma^s(\mathbf{q})\delta\Omega_\text{sw}
\mathcal{Q}^s\right. ,
\end{eqnarray}
where $\sigma^s(\mathbf{q})=\sigma_1e^{-i\mathbf{q}\cdot\mathbf{r}_1}$. 
With the help of matrix $\mathsf{O}^s$ Eq. (\ref{projSmols1}) 
can be re-written as
\begin{eqnarray}\label{projSmols2}
\Omega_\text{sw}^{s \text{irr}} &=& \mathcal{Q}^s\Omega_\text{sw}\mathcal{Q}^s +
\sum_{a,b}^2 \left. \pi_a^s(-\mathbf{q})\right> \left(\mathsf{O}^s\right)^{-1}_{ab}
\left<\pi_b^s(\mathbf{q})\right. +
\left. 2 \pi_3^s(-\mathbf{q})\right>\left(4\mathsf{O}^s\right)^{-1}_{33}
\left<2\pi_3^s(\mathbf{q})\right. 
\nonumber \\ &=& \mathcal{Q}^s\Omega_\text{sw}\mathcal{Q}^s +
\sum_{a,b} \left. \pi_a^s(-\mathbf{q})\right> \left(\mathsf{O}^s\right)^{-1}_{ab}(q)
\left<\pi_b^s(\mathbf{q})\right. .
\end{eqnarray}
We note that, like in the case of the collective dynamics irreducible operator,
one-particle reducible parts of $\mathcal{Q}^s\Omega_\text{sw}\mathcal{Q}^s$ are removed
separately for Brownian dynamics and particle size swaps. 

Next, we use
a formula analogous to Eq. (\ref{projident1}),
\begin{eqnarray}\label{projident3}
\frac{1}{z-\mathcal{Q}^s\Omega_\text{sw}\mathcal{Q}^s} 
&=& \frac{1}{z - \Omega_\text{sw}^{s \text{irr}}} 
\\ \nonumber && -  
\frac{1}{z - \Omega_\text{sw}^{s \text{irr}}}
\sum_{a,b} \left. \pi_a^s(-\mathbf{q})\right> \left(\mathsf{O}^s\right)^{-1}_{ab}(q)
\left<\pi_b^s(\mathbf{q})\right. 
\frac{1}{z-\mathcal{Q}^s\Omega_\text{sw}\mathcal{Q}^s}
%\nonumber \\
\end{eqnarray}
and we express $\mathsf{M}^{s \text{red}}$ in
terms of matrix $\mathsf{M}^{s \text{irr}}$ whose elements are functions 
evolving with the irreducible evolution operator,
\begin{eqnarray}\label{Ms33redirr}
\mathsf{M}^{s \text{red}}(q;z) =
\mathsf{M}^{s \text{irr}}(q;z) - 
\mathsf{M}^{s \text{irr}}(q;z) \left(\mathsf{O}^s\right)^{-1}(q)
\mathsf{M}^{s \text{red}}(q;z),
\end{eqnarray}
where 
\begin{eqnarray}\label{Ms33irr}
\mathsf{M}_{ab}^{s \text{irr}}(q;z) = 
\left<\pi_a^s(\mathbf{q}) \left(z-\Omega_{\text{sw}}^{s \text{irr}}\right)^{-1}
\pi_b^s(-\mathbf{q})\right>.
\end{eqnarray}

The combination of Eqs. (\ref{projident2}-\ref{Os33}), (\ref{Ms33}) 
and (\ref{Ms33redirr}-\ref{Ms33irr}) constitutes the memory function representation
for the tagged particle density correlation functions. 
We note that up to now, our approach was formally exact. On the other hand, 
it could be said that we have just re-wrote the problem of calculating
the tagged particle density correlation functions as a problem of
calculating the elements of matrix $\mathsf{M}^{s \text{irr}}$. Thus, the 
latter functions have become the main unknown objects of our theory.

\section{Factorization approximation}

To proceed, we need to introduce some approximations. Here, as in the earlier
study of the collective dynamics, we follow the spirit of the mode-coupling theory 
for glassy dynamics and the glass transition \cite{Goetzebook} and 
use a factorization approximation. 
More precisely, we use a sequence of three approximations 
\cite{Szamelswap,Goetzebook,SL}.  
In the present case these approximations need to be slightly modified to 
take into account the focus on the tagged particle dynamics.  

First, in expression (\ref{Ms33irr}) for the 
matrix elements of $\mathsf{M}^{s \text{irr}}$ 
we project functions $\pi_a^s$ onto the subspace spanned by the part of the product 
of the tagged particle density and the density of the other particles
orthogonal to the tagged particle density. The resulting expression reads,
\begin{eqnarray}\label{mct1}   
&& \mathsf{M}^{a \text{irr}}_{ab}(\mathbf{q}; t) \approx
\sum_{\alpha,...,\theta}\sum_{\mathbf{k}_1,...,\mathbf{k}_x}
\left<\pi_a^s(\mathbf{q}) 
n_\alpha^s(-\mathbf{k}_1)n_\beta(-\mathbf{k}_2)\right> 
\nonumber \\ && \times
\left<n_\alpha^s(\mathbf{k}_1)n_\beta(\mathbf{k}_2)\mathcal{Q}^s
n_\gamma^s(-\mathbf{k}_3)n_\delta(-\mathbf{k}_4)\right> ^{-1}
\nonumber \\ && \times
\left<n_\gamma^s(\mathbf{k}_3)n_\delta(\mathbf{k}_4)
\exp(\Omega_\text{sw}^{s \text{irr}}t)
n_\epsilon^s(-\mathbf{k}_5)n_\zeta(-\mathbf{k}_6)\right>
\nonumber \\ && \times 
\left<n_\epsilon^s(\mathbf{k}_5)n_\zeta(\mathbf{k}_6)\mathcal{Q}^s
n_\eta^s(-\mathbf{k}_7)n_\theta(-\mathbf{k}_8)\right>^{-1}
\left< n_\eta^s(\mathbf{k}_7)n_\theta(\mathbf{k}_8)\pi_b^s(-\mathbf{q})
\right>.
\end{eqnarray}
In the case of Brownian dynamics with pairwise additive interactions the
first approximation is exact. In the present case of Brownian dynamics
with particle size swaps, Eq. (\ref{mct1}) constitutes an approximation, due
to the fact that factors $w_i$ are typically not pairwise additive.  

Second, we factorize four-point dynamic 
correlation functions while replacing the tagged particle irreducible evolution 
operator by the original un-projected evolution operator, 
\begin{eqnarray}\label{mct2}  
&& \left<n_\gamma^s(\mathbf{k}_3)n_\delta(\mathbf{k}_4)\mathcal{Q}^s 
\exp(\Omega_\text{sw}^{s \text{irr}} t) 
\mathcal{Q}^s n_\epsilon^s(-\mathbf{k}_5)n_\zeta(-\mathbf{k}_6)\right>
\nonumber \\ &\approx&
\left<n_\gamma^s(\mathbf{k}_3) \exp(\Omega_\text{sw} t) 
n_\epsilon^s(-\mathbf{k}_5) \right>
\left<n_\delta(\mathbf{k}_4) \exp(\Omega_\text{sw} t) n_\zeta(-\mathbf{k}_6) \right>
\nonumber \\ & \equiv &
\left<n_\gamma^s(\mathbf{k}_3;t) n_\epsilon^s(-\mathbf{k}_5) \right>
\left<n_\delta(\mathbf{k}_4;t) n_\zeta(-\mathbf{k}_6) \right>.
\end{eqnarray}
We note that Eq. (\ref{mct2}) is the crucial approximation of our approach.

Third, we use some additional approximations for static correlation functions. 
We factorize normalization factors
$\left<n_\alpha^s(\mathbf{k}_1)n_\beta(\mathbf{k}_2)\mathcal{Q}^s
n_\gamma^s(-\mathbf{k}_3)n_\delta(-\mathbf{k}_4)\right> ^{-1}$,
\begin{eqnarray}\label{mct3}
&& \left<n_\alpha^s(\mathbf{k}_1)n_\beta(\mathbf{k}_2)\mathcal{Q}^s
n_\gamma^s(-\mathbf{k}_3)n_\delta(-\mathbf{k}_4)\right>^{-1}
\nonumber \\ && 
\approx 
\left< n_\alpha^s(\mathbf{k}_1)n_\gamma^s(-\mathbf{k}_3) \right>^{-1}
\left< n_\beta(\mathbf{k}_2)n_\delta(-\mathbf{k}_4) \right>^{-1},
\end{eqnarray}
and we note that $\left< n_\alpha^s(\mathbf{k}_1)n_\gamma^s(-\mathbf{k}_3) \right>^{-1}
=x_\alpha^{-1}\delta_{\mathbf{k}_1\mathbf{k}_3}\delta_{\alpha\gamma}$ and
$\left< n_\beta(\mathbf{k}_2)n_\delta(-\mathbf{k}_4) \right>^{-1} = 
\delta_{\mathbf{k}_2\mathbf{k}_4}S_{\beta\delta}^{-1}(k_2)$, with $S_{\beta\delta}(k)$
being the partial static structure factor. We note that the vertices
originating from Brownian dynamics,
$\left< n_\eta^s(\mathbf{k}_7)n_\theta(\mathbf{k}_8)\pi_b^s(-\mathbf{q})\right>$,
$b=1,2$, can be calculated exactly, 
\begin{eqnarray}\label{mct4}
\left< n_\eta^s(\mathbf{k}_7)n_\theta(\mathbf{k}_8)\pi_b^s(-\mathbf{q})
\right> = \frac{nx_b}{\sqrt{N}}
\delta_{\mathbf{k}_7+\mathbf{k}_8,\mathbf{q}} \mathbf{k}_8\cdot \mathbf{q}
\delta_{\eta b} \sum_\gamma S_{\theta\gamma}(k_8)c_{\gamma b}(k_8)
\;\;\;\;\; b=1,2.
\nonumber \\
\end{eqnarray}
The calculation of the vertex originating from particle size swaps,
$\left< n_\eta^s(\mathbf{k}_7)n_\theta(\mathbf{k}_8)\pi_3^s(-\mathbf{q})\right>$,
is a bit more involved. Using the mixture version of the convolution 
approximation we initially obtain the following expression
\begin{eqnarray}\label{mct5}
&& \left< n_\eta^s(\mathbf{k}_7)n_\theta(\mathbf{k}_8)\pi_3^s(-\mathbf{q})\right> 
= -\left(2\tau_\text{sw}\sqrt{N}\right)^{-1}
\delta_{\mathbf{k}_7+\mathbf{k}_8,\mathbf{q}} 
\left(\delta_{\eta A}-\delta_{\eta B}\right)
\nonumber \\ && \times 
\left[R_\theta(k_8)-\frac{1}{2}\left(Q\left(\delta_{\theta A}-\delta_{\theta B}\right)
+\left<w_1\right>\right)
-\left< w_1 \right>x_\eta^{-1}
\left(S_{\eta\theta}(k_8)-x_\theta\delta_{\eta\theta}\right)\right]
\end{eqnarray}
where 
\begin{eqnarray}\label{RQa}
R_\theta(q) = \left< n_{\theta}(\mathbf{q})
w_1 e^{i\mathbf{q}\cdot\mathbf{r}_1}\right> 
\;\;\;\text{ and }\;\;\; Q = \left< \sigma_1 w_1 \right>
\end{eqnarray}
To simplify expression (\ref{mct5}) we factor out 
$\left<w_1\right>$ from $R_\theta$ and $Q$,
\begin{eqnarray}\label{RQb}
R_\theta(q) \approx \left<w_1 \right>
\sum_{\lambda} S_{\theta\lambda}(q)  
\;\;\;\text{ and } \;\;\; Q \approx  
\left< w_1 \right> \left< \sigma_1 \right>
= \left< w_1 \right> \left(x_A-x_B\right)
\end{eqnarray}
Using Eq. (\ref{RQb}) in Eq. (\ref{mct5}) we obtain 
\begin{eqnarray}
&& \left< n_\eta^s(\mathbf{k}_7)n_\theta(\mathbf{k}_8)\pi_3^s(-\mathbf{q})\right> 
= -\left(2\tau_\text{sw}\sqrt{N}\right)^{-1}\delta_{\mathbf{k}_7+\mathbf{k}_8,\mathbf{q}}
\left(\delta_{\eta A}-\delta_{\eta B}\right)
\nonumber \\ && \times 
\left<w_1\right> \left[\sum_{\lambda} S_{\theta\lambda}(k_8)
-x_A\delta_{\theta A} - x_B\delta_{\theta B}
- x_\eta^{-1}\left(S_{\eta\theta}(k_8)-x_\theta\delta_{\eta\theta}\right)\right].
\end{eqnarray}

The final result is a set of approximate expressions for the
elements of matrix $\mathsf{M}^{s \text{irr}}$ in terms of integrals of products
of the tagged particle and collective density correlation functions. Here we list 
these expressions for the equimolar mixture, $x_A=x_B=0.5$, which was considered
in Ref. \cite{Szamelswap}, in the thermodynamic limit, 
$N\to\infty, V\to\infty, N/V=n=\text{const.}$ 
The expressions for $\mathsf{M}^{s \text{irr}}_{ab}$,
$a,b=1,2$ are the same as those derived in the standard mode-coupling theory,
\begin{eqnarray}\label{Ms33mctBD}
\mathsf{M}^{s\text{irr}}_{ab}(\mathbf{q}; t) & \approx & nD_0^2 
\int \frac{d\mathbf{k}}{(2\pi)^3} \sum_{\beta\delta} \mathbf{q}\cdot\mathbf{k}
c_{a\beta}(k) F^s_{ab}(|\mathbf{q}-\mathbf{k}|;t)
\nonumber \\ && \times
F_{\beta\delta}(k;t)c_{\delta b}(k)
\mathbf{q}\cdot\mathbf{k}.
\end{eqnarray}
In Eq. (\ref{Ms33mctBD}) and in the following equations, 
$c_{\alpha 1}\equiv c_{\alpha A}$ and $c_{\alpha 2}\equiv c_{\alpha B}$.
The remaining matrix elements originate from particle size swaps,
\begin{eqnarray}\label{Ms33mctmixed}
\mathsf{M}^{s \text{irr}}_{a3}(\mathbf{q}; t) & \approx & 
-nD_0 \frac{\left< w_1 \right>}{2\tau_\text{sw}}
\int \frac{d\mathbf{k}}{(2\pi)^3} \sum_{\beta\gamma\delta} \mathbf{q}\cdot\mathbf{k}
c_{a\beta}(k) F^s_{a\gamma}(|\mathbf{q}-\mathbf{k}|;t)
\nonumber \\ && \times
F_{\beta\delta}(k;t)\left(c_{\delta A}(k)-c_{\delta B}(k)\right) = 
\mathsf{M}^{s \text{irr}}_{3a}(\mathbf{q}; t) \;\;\;\;\; a=1,2,
\end{eqnarray}
\begin{eqnarray}\label{Ms33mctswap}
\mathsf{M}^{s \text{irr}}_{33}(\mathbf{q}; t) & \approx & 
n\frac{\left< w_1 \right>^2}{\left(2\tau_\text{sw}\right)^2}
\int \frac{d\mathbf{k}}{(2\pi)^3} \sum_{\alpha\beta\gamma\delta}
\left(c_{A\beta}(k)-c_{B\beta}(k)\right)
F^s_{\alpha\gamma}(|\mathbf{q}-\mathbf{k}|;t)
\nonumber \\ && \times
F_{\beta\delta}(k;t)\left(c_{\delta A}(k)-c_{\delta B}(k)\right).
\end{eqnarray}

The exact memory function representation defined through Eqs. 
(\ref{projident2}-\ref{Os33}), (\ref{Ms33}) and (\ref{Ms33redirr}-\ref{Ms33irr}), 
combined with approximate form of the elements of matrix $\mathsf{M}^{s \text{irr}}$ 
in terms of integrals of products of the tagged particle and collective density 
correlation functions, Eqs. (\ref{Ms33mctBD}-\ref{Ms33mctswap}), constitute our 
mode-coupling theory for the single-particle motion. Numerical solution of these
time-dependent equations looks quite a bit more difficult than the solution
of the standard mode-coupling equations for mixtures. The reason for this supposition
is that in the present case the relation between the memory function and the
correlation functions is more complicated.

We note that, as in the standard mode-coupling theory, the necessary inputs into
the theory for the single-particle dynamics are collective intermediate scattering
functions, which in the present case should be obtained from the theory for
the collective dynamics with particle size swaps developed in Ref. \cite{Szamelswap}.

\section{Numerical results for the arrested phase}

First, let us recall the analysis 
presented in Ref. \cite{Szamelswap}. We started from equations for the 
time-dependent collective density correlation functions derived 
and we assumed that as the density increases and/or 
the temperature decreases, collective density correlations develop plateaus, 
and that at the transition nonzero long-time limits $F_{\alpha\beta}(q; \infty)$ appear 
discontinuously. These assumptions allowed us to derive a set of self-consistent 
equations for $F_{\alpha\beta}(q; \infty)$. We analyzed these self-consistent equations 
numerically for an equimolar binary hard-sphere mixture using as the static input 
equilibrium correlation functions obtained from the Percus-Yevick (PY) closure 
\cite{PY1,PY2,Voigtmannthesis}. We obtained a fluid-glass phase diagram. 
We found that for a system with particle size swaps the volume fraction 
at the dynamic glass transition increases with increasing particle diameter ratio 
up to the ratio of about 1.2, where the relative increase is about 4\%, and then
saturates. In agreement with Ref. \cite{GoetzeVoigtmann}, the location of the 
dynamic glass transition for an equimolar mixture without particle size swaps 
depends very weakly on the ratio of particle diameters for the ratio smaller than 1.5.

Here, we focus on the non-decaying parts of the tagged particle density
correlations $F_{\alpha\beta}^s(q; \infty)$. Their appearance signals the arrest of
the single-particle motion. We recall that in the standard mode-coupling theory
the dynamic glass transition, \textit{i.e.} the appearance of the frozen-in
collective density fluctuations, coincides with the arrest of the single-particle
motion. We note that this prediction is non-trivial. Generally speaking, if one
considers single-particle motion in a fluid with some frozen-in disorder, the
arrest of the single-particle motion is only predicted if the disorder is strong enough.
It follows from an explicit calculation that the frozen-in density fluctuations 
predicted by the standard mode-coupling theory at the dynamic glass transition 
are strong enough to arrest the single-particle motion (under the assumption that 
the latter process is described by the mode-coupling theory for the single-particle 
motion). 

First, we derive the self-consistent equations for the non-decaying parts of the 
tagged particle density correlations $F_{\alpha\beta}^s(q; \infty)$. We start 
from the combination of Eqs. (\ref{projident2}-\ref{Os33}), (\ref{Ms33}), 
(\ref{Ms33redirr}-\ref{Ms33irr}) and (\ref{Ms33mctBD}-\ref{Ms33mctswap})
and obtain
\begin{eqnarray}\label{arrest}
\lefteqn{ 
F_{\alpha\beta}^s(q;\infty) - x_\alpha \delta_{\alpha\beta}(q) = }
\nonumber \\ && 
- \sum_{\gamma\delta}
\mathsf{v}^{\text{T}}_\alpha \mathsf{O}^s(q) 
\left[\mathsf{M}^{s \text{irr}}(q,\infty)\right]^{-1} \mathsf{O}^s(q) \mathsf{v}_\gamma
x_\gamma^{-1} \delta_{\gamma\delta} F_{\delta\beta}^s(q;\infty),
\end{eqnarray}
where $\mathsf{M}^{s \text{irr}}(q,\infty)$ is given by 
Eqs. (\ref{Ms33mctBD}-\ref{Ms33mctswap}) with non-zero long-time limits 
$F_{\alpha\beta}(q;\infty)$ and $F_{\alpha\beta}^s(q;\infty)$ 
substituted at their right-hand-sides.

We solved self-consistent equations (\ref{arrest}) numerically \cite{details}, 
using as the static 
input equilibrium correlation functions obtained from the Percus-Yevick (PY) closure 
\cite{PY1,PY2,Voigtmannthesis} and using previously obtained non-decaying parts
of the collective density fluctuations $F_{\alpha\beta}(q;\infty)$. We found that,
as is the case in the standard mode-coupling theory, the single-particle
motion becomes arrested at the location of the dynamic glass transition. 

In the first two sets of figures below, Figs. \ref{fig:sw-nsw}-\ref{fig:nsw-norm}, 
we compare the results obtained from the present theory for 
an equimolar binary hard-sphere mixture with particle size swaps with the 
results of the standard mode-coupling theory for the same system with Brownian dynamics
only. In both cases we present results at the dynamic glass transition, respectively 
the transition predicted by the present theory and that predicted by the standard
mode-coupling theory. In all the figures we present results for the 
diameter ratio at which the difference between these transitions is approximately
the largest, $d_L/d_S=1.2$.

In Figs. \ref{fig:sw-nsw} we compare the non-decaying parts of the 
collective density correlations for the larger particles, $F_{AA}(q; \infty)$,
with the corresponding partial structure factors $S_{AA}(q)$ at the same
volume fraction. We see that in both cases the wavevector dependence of 
$F_{AA}(q; \infty)$ follows that of $S_{AA}(q)$. The main difference between 
the two results is seen in the zero wavevector limit. 
The present theory predicts $\lim_{q\to 0} F_{AA}(q; \infty)$ is significantly smaller 
than $\lim_{q\to 0} S_{AA}(q)$, in contrast with the 
standard mode-coupling binary mixture result, 
$\lim_{q\to 0} F_{AA}(q; \infty) \approx \lim_{q\to 0} S_{AA}(q)$. 
This qualitative difference originates from the particle size swaps.

\begin{figure}
\includegraphics[scale=0.3]{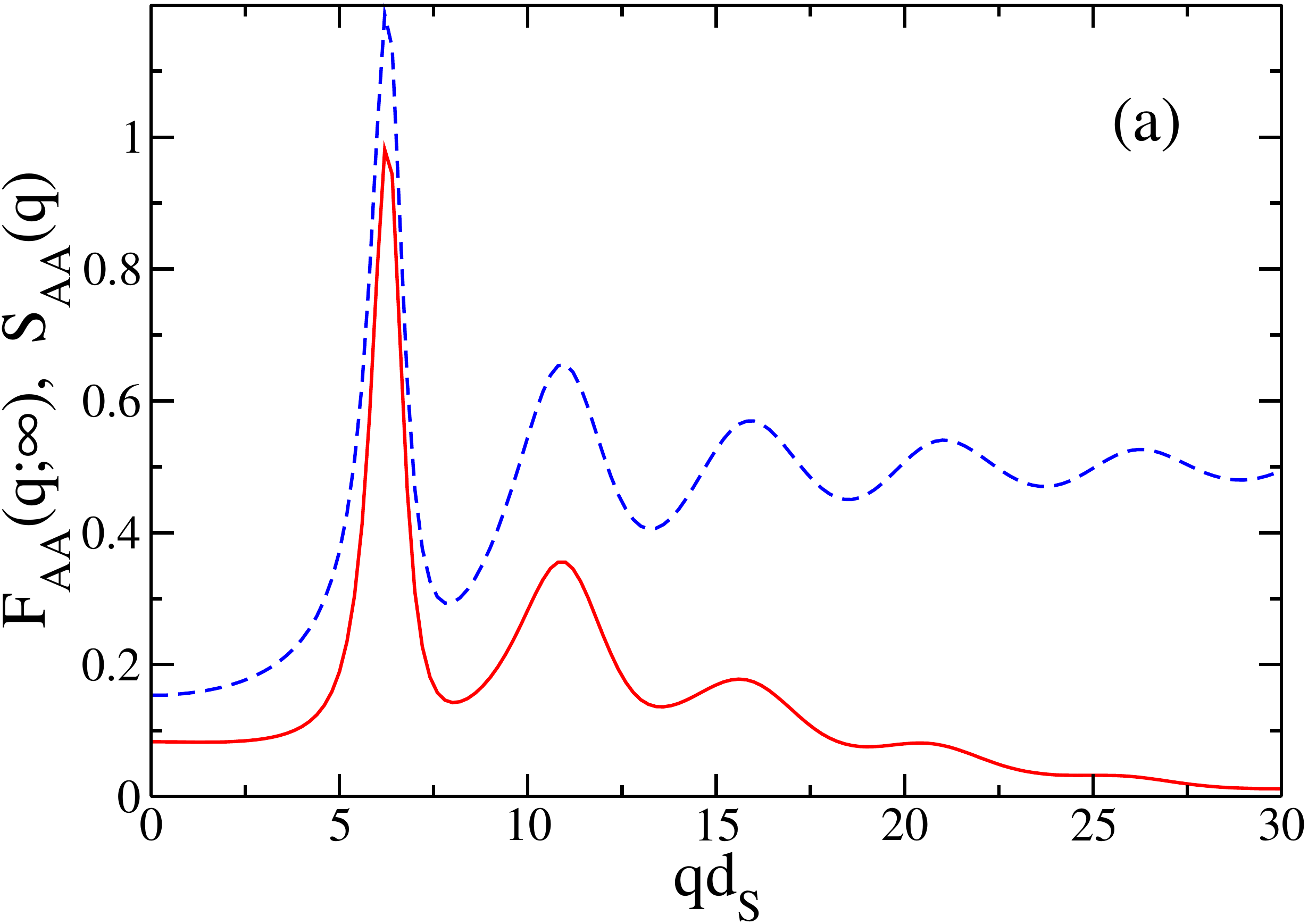} \hfill 
\includegraphics[scale=0.3]{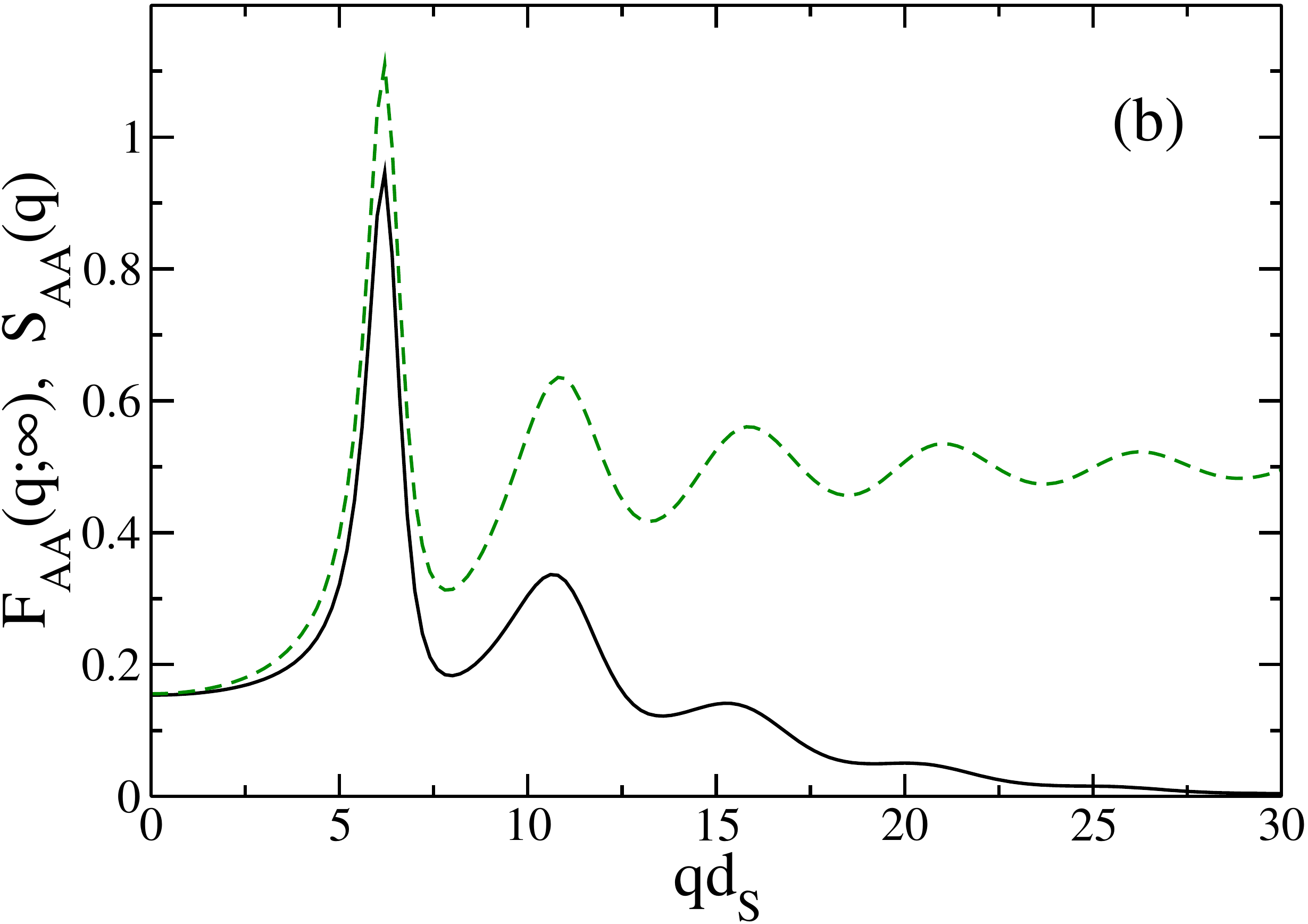}
\caption{\label{fig:sw-nsw} Non-decaying part of the large particles collective density 
correlation function, $F_{AA}(q; \infty)$ (solid line) compared to the large particles 
partial structure factor, $S_{AA}(q)$ (dashed line). 
Left panel (a): with particle size swaps; right panel (b): without swaps.}
\end{figure}

In the next set of figures we compare the non-decaying parts of the reduced, diagonal 
collective density correlations $F_{\alpha\alpha}(q; \infty)/S_{\alpha\alpha}(q)$, 
$\alpha=A,B$ and tagged particle density fluctuations, $F_{\alpha\alpha}^s(q; \infty)$,
obtained from the present theory, Figs. \ref{fig:sw-norm}a-b, and from the
standard mode-coupling theory, Figs. \ref{fig:nsw-norm}a-b. Now we clearly see the 
difference of the small wavevector behavior of the frozen-in collective density
correlations discussed above. In addition, we see the qualitative difference 
between the non-decaying parts of the diagonal tagged particle 
density correlations $F_{\alpha\alpha}^s(q; \infty)$. The present theory predicts
$\lim_{q\to 0} F_{\alpha\alpha}^s(q; \infty)/x_\alpha < 1$ whereas within the standard 
mode-coupling approach the conservation of the number of the particles of each 
type leads to $\lim_{q\to 0} F_{\alpha}^s(q; \infty)/x_\alpha = 1$.

\begin{figure}
\includegraphics[scale=0.3]{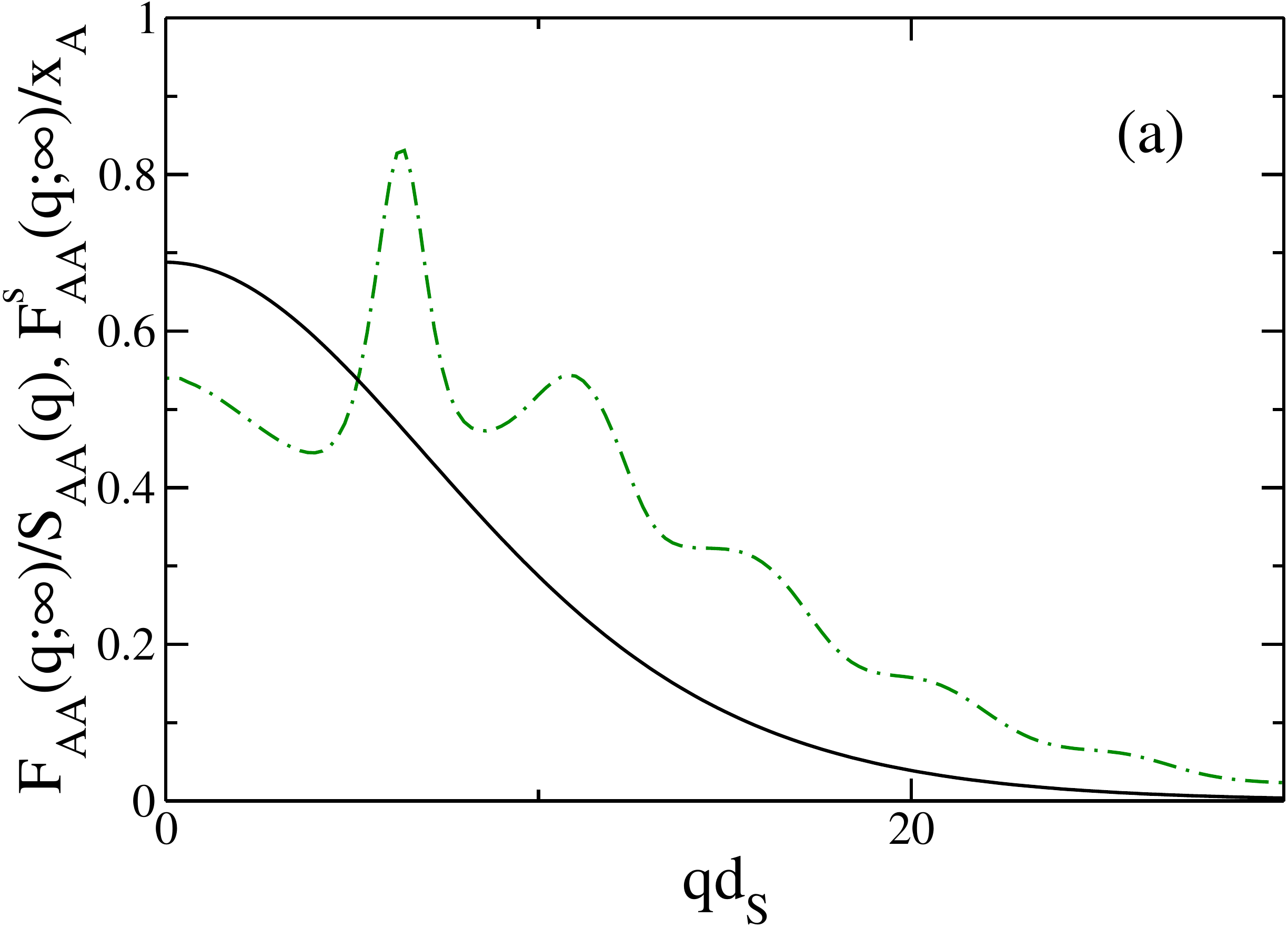} \hfill 
\includegraphics[scale=0.3]{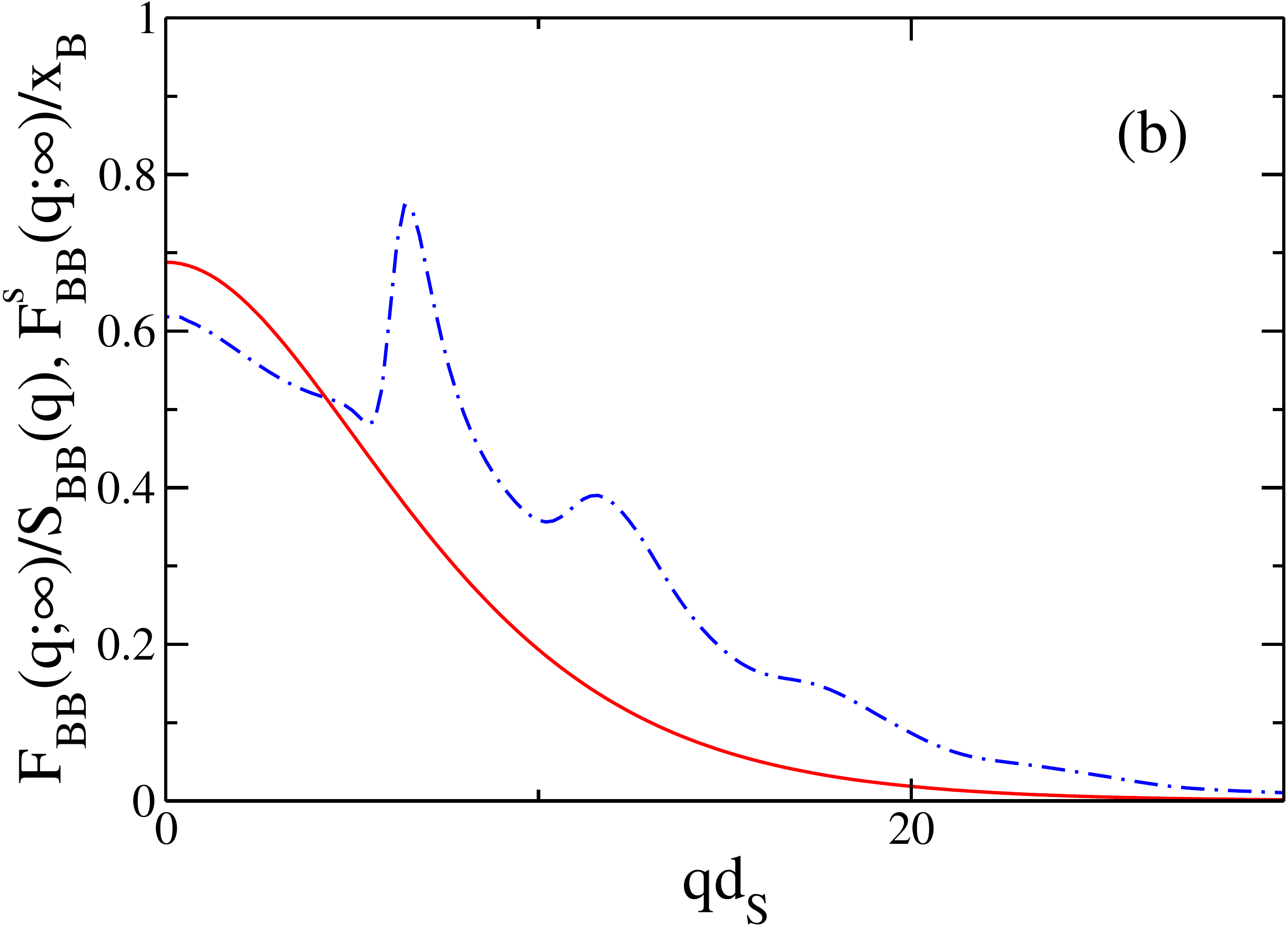}
\caption{\label{fig:sw-norm} Particle size swap dynamics. 
Non-decaying part of the tagged particle density  
correlation function, $F_{\alpha\alpha}^s(q; \infty)/x_\alpha$ (solid line), 
compared to the corresponding normalized collective density correlation function,
$F_{\alpha\alpha}^s(q; \infty)/S_{\alpha\alpha}(q)$ (dashed line). 
Left panel (a): $\alpha=A$; right panel (b): $\alpha=B$.}
\end{figure}

\begin{figure}
\includegraphics[scale=0.3]{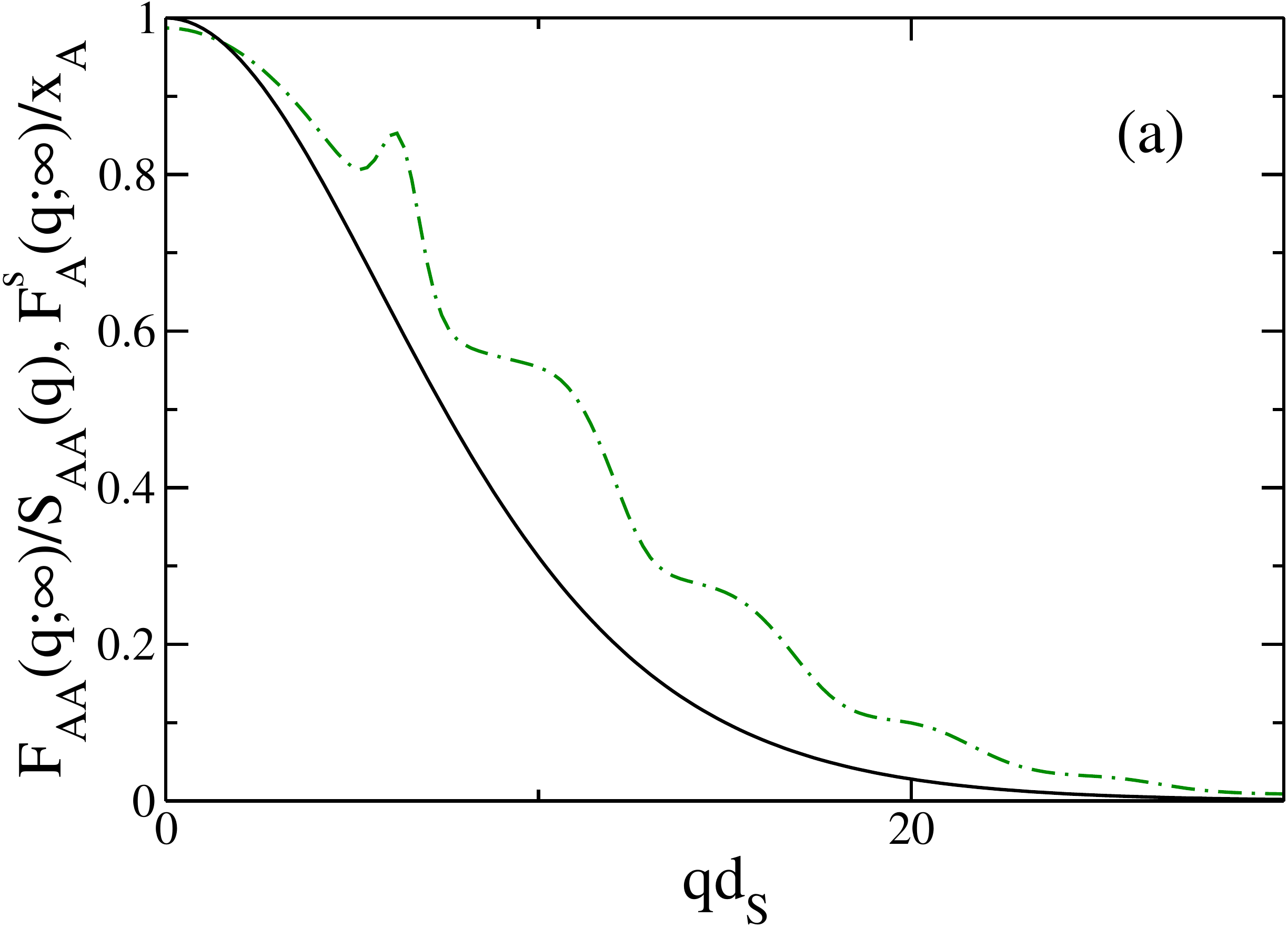} \hfill 
\includegraphics[scale=0.3]{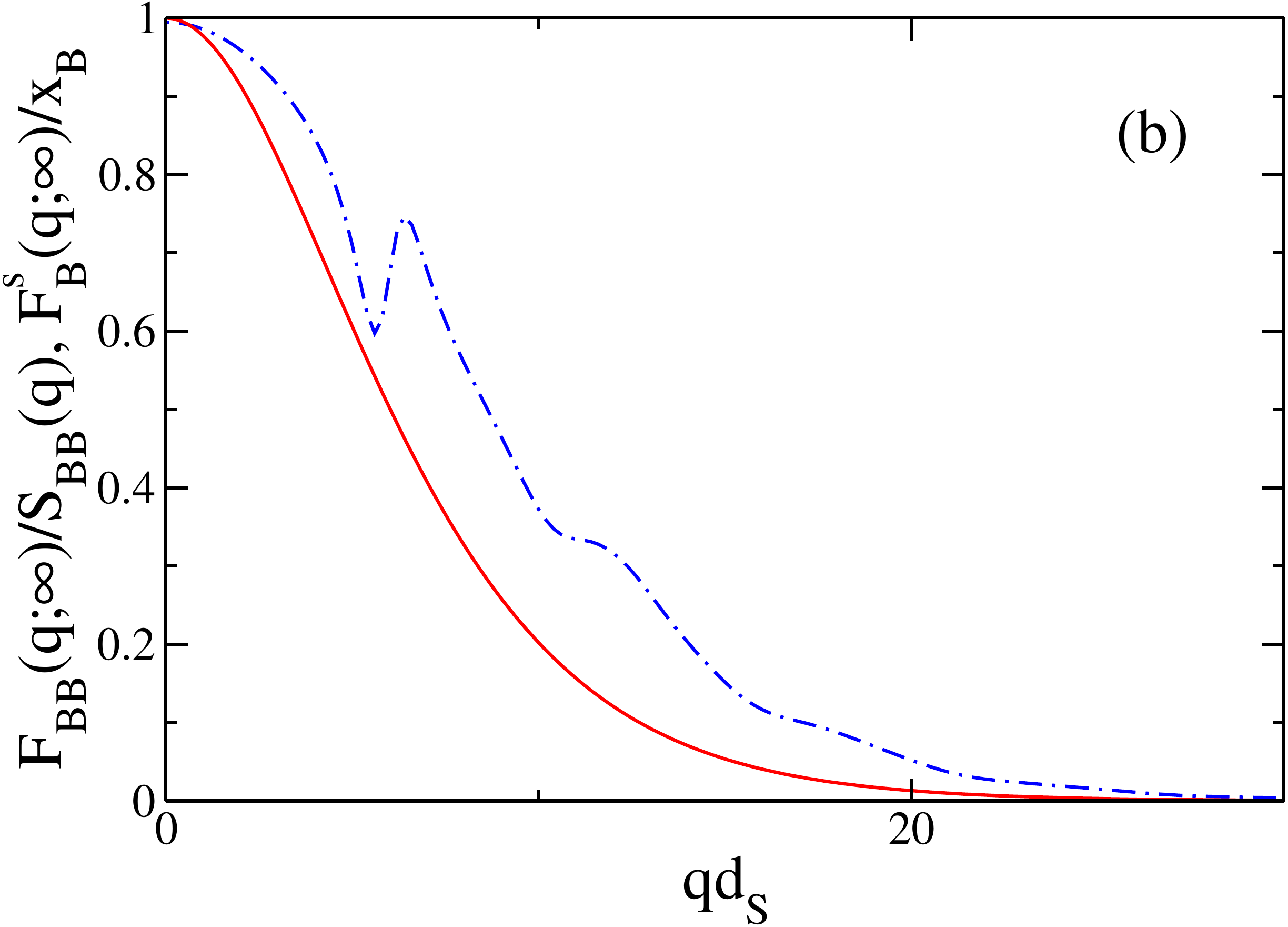}
\caption{\label{fig:nsw-norm} Dynamics without particle size swaps.  
Non-decaying part of the tagged particle density  
correlation function, $F_{\alpha}^s(q; \infty)/x_\alpha$ (solid line), compared to the 
corresponding normalized collective density correlation function,
$F_{\alpha\alpha}(q; \infty)/S_{\alpha\alpha}(q)$ (dashed line). 
Left panel (a): $\alpha=A$; right panel (b): $\alpha=B$.}
\end{figure}

Following the spirit of Ikeda \textit{et al.} \cite{IkedaZamponiIkeda,IkedaZamponi} we can
use the small wavevector limits of the frozen-in tagged particle 
density correlations $F_{\alpha\alpha}^s(q; \infty)$ to define an order parameter
for particle size swaps in the glass,
\begin{eqnarray}\label{qqdef}
q= \lim_{q\to 0} 
\left[F_{AA}^s(q; \infty)-F_{AB}^s(q; \infty)-F_{BA}^s(q; \infty)+F_{BB}^s(q; \infty)
\right],
\end{eqnarray}
where use the same symbol $q$ as in Refs. \cite{IkedaZamponiIkeda,IkedaZamponi}
but caution the reader no to confuse this quantity with the wavevector $q$.
The zero value of the order parameter, $q=0$ indicates the complete lack 
of particle type correlations in the glass whereas the maximum value, $q=1$,
indicates the absence of the particle size swaps in the glass.

\begin{figure}
\includegraphics[scale=0.3]{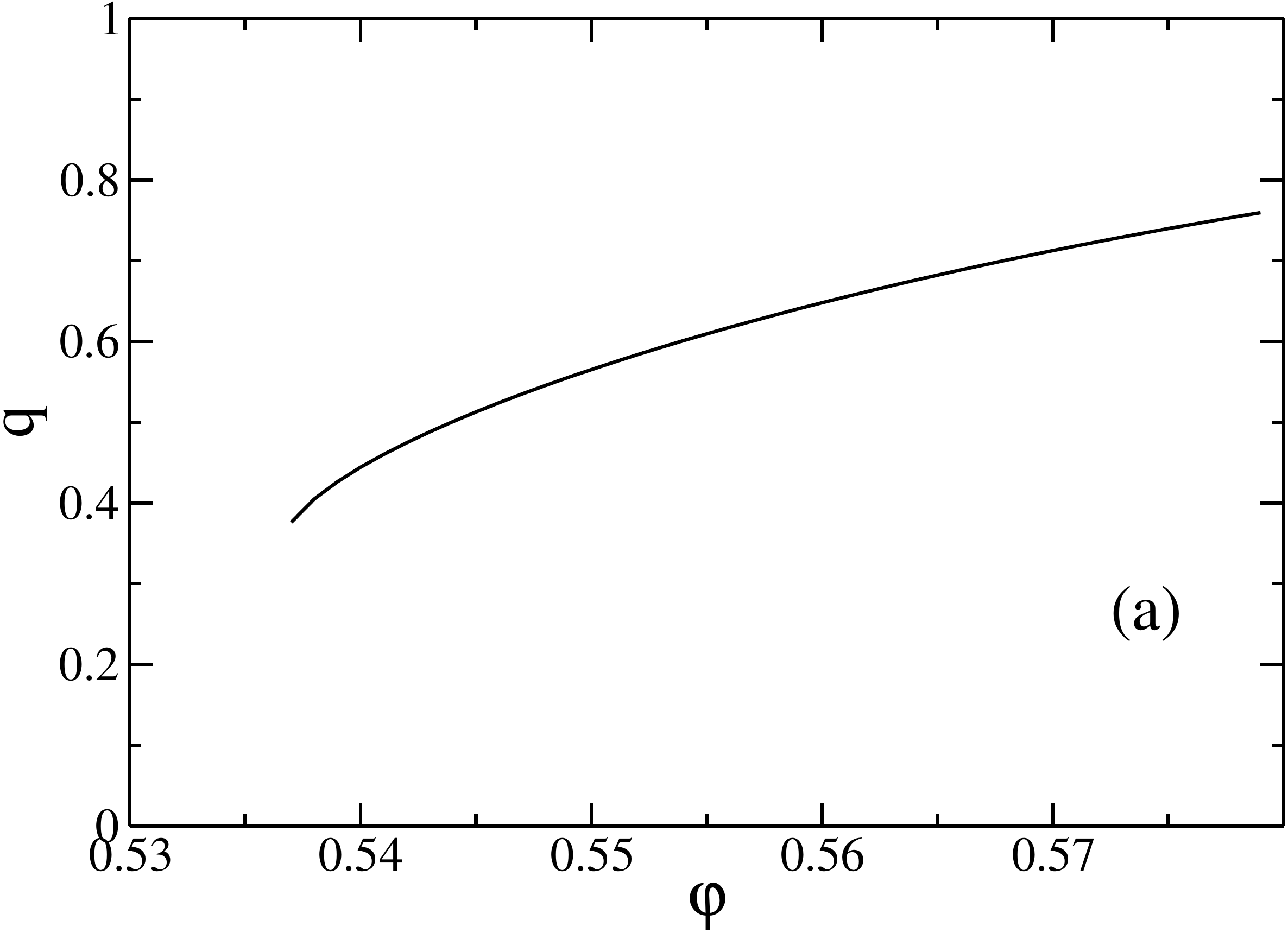} \hfill 
\includegraphics[scale=0.3]{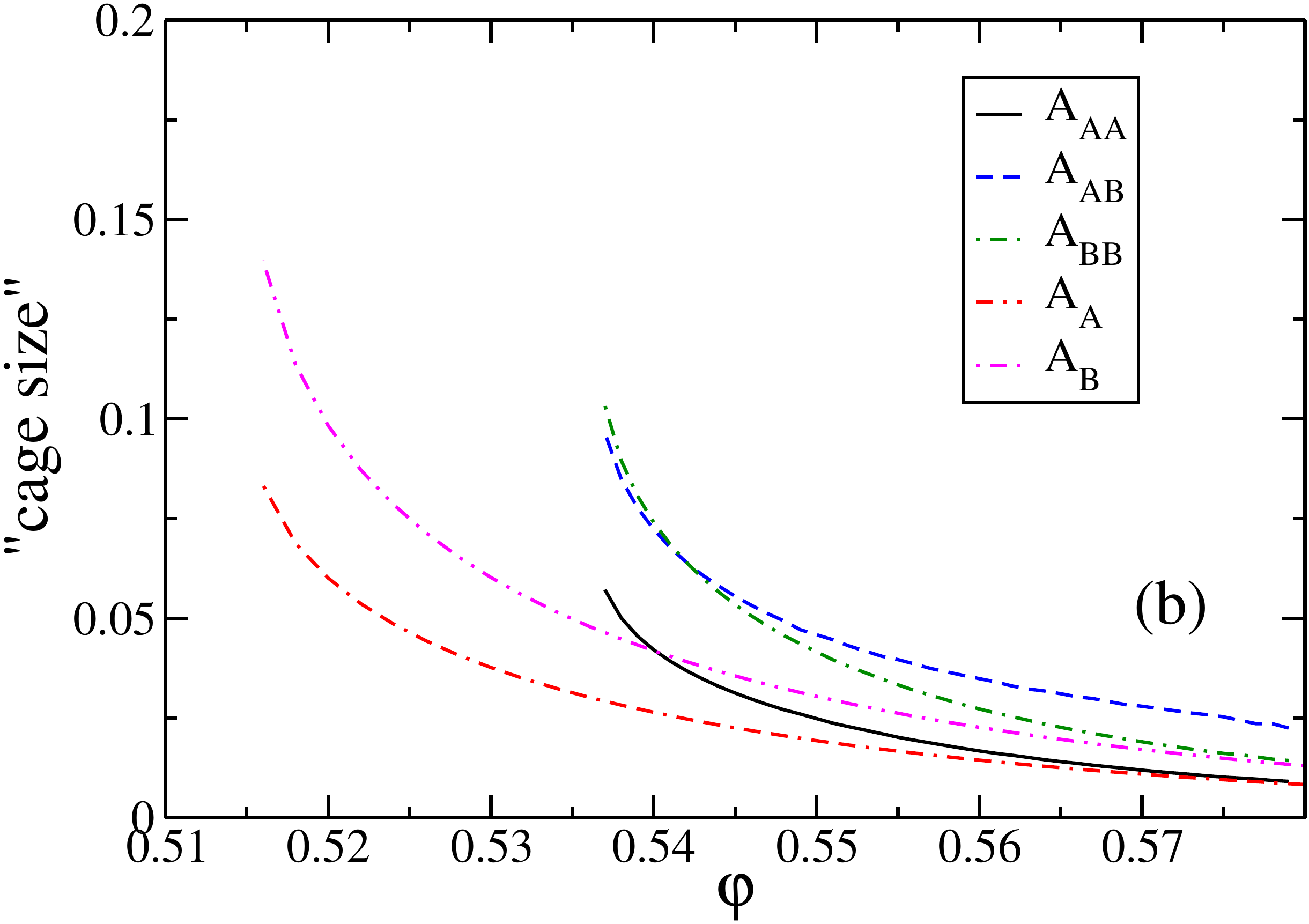}
\caption{\label{fig:qq-radii} Left panel (a): order parameter for particle size swaps,
$q$, as a function of the volume fraction, $\varphi$.
Right panel (b): long-time limit of the mean square displacement (``cage size'') 
for the system with swap dynamics ($A_{\alpha\beta}, \alpha,\beta=A,B$) and 
for the system without swaps ($A_{\alpha}, \alpha=A,B$).}
\end{figure}

In Fig. \ref{fig:qq-radii}a we show the volume fraction dependence of the 
particle size swaps order parameter for the 
diameter ratio at which the difference between these transitions is approximately
the largest, $d_L/d_S=1.2$. On physical grounds one would expect the particle size 
swaps to become less probable with increasing volume fraction. The present theory
confirms this expectation: we observe $q$ increasing monotonically starting 
from a nonzero value $q\approx 0.376$ at the dynamic glass transition.
Qualitatively, the volume fraction dependence of the particle size swaps order parameter
predicted by the present theory agrees with the prediction of the replica approach
of Ikeda and Zamponi \cite{IkedaZamponi}. 

Within the standard mode-coupling theory, 
the small wavevector behavior of the frozen-in tagged particle 
density correlations $F_{\alpha}^s(q; \infty)$ allows one to define the 
long-time limit of the mean square displacement in the glass, which can
be interpreted as a quantification of the ``cage size'',
\begin{eqnarray}\label{msdnswp}
F_\alpha^s(q;\infty) = x_\alpha\left( 1 - \frac{1}{2}A_\alpha q^2/2
+ ... \right) \; \text{ where } \; A_\alpha=
\lim_{t\to\infty} \left<|\mathbf{r}_1^\alpha(t)-\mathbf{r}_1^\alpha(0)|^2\right>.
\nonumber \\
\end{eqnarray} 

In the present case of a system with particle size swaps we generalize this 
definition as
\begin{eqnarray}\label{msdswp}
&& F_{\alpha\beta}^s(q;\infty) = \sqrt{x_\alpha x_\beta} 
\left( 1 - \frac{1}{2}A_{\alpha\beta} q^2/2
+ ... \right) 
\nonumber \\ && 
\;\;\;\;\;\;\;\;\;\; \text{ where } \;\;\;A_{\alpha\beta}=
\lim_{t\to\infty} \left<|\mathbf{r}_1^\alpha(t)-\mathbf{r}_1^\beta(0)|^2\right>.
\end{eqnarray} 

In Fig. \ref{fig:qq-radii}b we show the volume fraction dependence of the
$A_{\alpha\beta}$ and $A_\alpha$ for the 
diameter ratio at which the difference between these transitions is approximately
the largest, $d_L/d_S=1.2$. As expected on physical grounds, all the quantities 
characterizing ``cage size'' decrease with increasing volume fraction. For 
a given volume fraction, the ``cage size'' for the system with particle size swaps is
larger than that for the system without swaps. This is consistent 
with simulational results of Ninarello \textit{et al} \cite{NinarelloPRX}. 
Furthermore, as expected on physical grounds, the cage size depends on the 
particle type and is smaller for the larger particles. 
Again, qualitatively, the volume fraction dependence of the ``cage size'' 
predicted by the present theory agrees with the prediction of the replica approach
of Ikeda and Zamponi \cite{IkedaZamponi}. However, the important difference is
that the latter theory uses only a single cage size parameter for both types of the
particles, both in the 
system with particle size swaps and in the system without swaps.

\section{Discussion}

We presented here a theory for the single-particle motion in binary
mixtures with particle size swaps. The present theory is an extension of the
previously developed theory for the collective dynamics and the dynamic glass
transition in binary mixtures with particle size swaps. Like the latter approach,
the present theory is based on combination of a formally exact memory function
representation of the tagged particle density correlation function and a 
factorization approximation for a four-point correlation function evolving with 
the irreducible dynamics. Thus, it belongs to the class of mode-coupling-like
theories. 

The present theory predicts that the single-particle motion becomes localized
at the dynamic glass transition, like in the standard mode-coupling theory for
systems with local dynamics only. The properties of the localized state are 
qualitatively similar to those of the localized state of a system with local 
dynamics only. The main difference is that at the same volume fraction the 
``cage size'' is predicted to be larger in a system with particle size swaps 
than in the corresponding system without swaps. 

The properties of the localized state predicted by the present theory are 
qualitatively similar to those predicted by the replica approach \cite{IkedaZamponi}.
The main difference is that in our approach different ``cage sizes'' are 
obtained from the (approximate) calculation whereas the replica
\textit{ansatz} assumes that there is a single ``cage size'' parameter. 

While solving numerically the time-dependent equations of motion for either
collective or tagged particle density correlations seems difficult, it might be 
possible to re-do the standard mode-coupling asymptotic analysis and
to check whether the power law approach to and departure from intermediate
time plateaus are still present, and to check whether there is an analogue
of the separation parameter determining the corresponding power laws exponents.
This interesting problem is left for a future study.

\section*{Acknowledgments}

I thank Th. Voigtmann for a reference to explicit expressions for binary hard
sphere mixture PY structure factors and E. Flenner for
comments on the manuscript.
I gratefully acknowledge partial support of NSF Grants No.~DMR-1608086
and No.~CHE-1800282.

\section*{}


\begin{thebibliography}{99}
\bibitem{BerthierBiroliRMP} L. Berthier and G. Biroli,
Rev. Mod. Phys. \textbf{83}, 587 (2011).
\bibitem{BerthierPNAS} L. Berthier, P. Charbonneau, D. Coslovich, A. Ninarello,
M. Ozawa, and S. Yaida,
Proc. Natl. Acad. Sci. USA \textbf{114}, 11356 (2017).
\bibitem{Grigera} T. Grigera and G. Parisi, Phys. Rev. E \textbf{63}, 045102(R) (2001).
\bibitem{Flenner2006} See, \textit{e.g.}, E. Flenner and G. Szamel, 
Phys. Rev. E \textbf{73}, 061505 (2006).
\bibitem{BerthierCNO} L. Berthier, D. Coslovich, A. Ninarello, and M. Ozawa,
Phys. Rev. Lett. \textbf{116}, 238002 (2016).
\bibitem{NinarelloPRX} A. Ninarello, L. Berthier, and D. Coslovich,
Phys. Rev. X \textbf{7}, 021039 (2017).
\bibitem{OzawaPNAS} M. Ozawa, L. Berthier, G. Biroli, A. Rosso and G. Tarjus,
PNAS \textbf{115}, 6656 (2018).
\bibitem{WangNatComm} L. Wang, A. Ninarello, P. Guan, L. Berthier, G. Szamel and
E. Flenner,
Nature Communications \textbf{10}, 26 (2019).
\bibitem{Berthier2d} L. Berthier, P. Charbonneau, A. Ninarello, M. Ozawa
and S. Yaida, arXiv:1805.09035.
\bibitem{WyartCates} M. Wyart and M.E. Cates,
Phys. Rev. Lett. \textbf{119}, 195501 (2017).
\bibitem{BBBT} L. Berthier, G. Biroli, J.-P. Bouchaud, and G. Tarjus,
arXiv:1805.12378.
\bibitem{IkedaZamponiIkeda} H. Ikeda, F. Zamponi, and A. Ikeda,
J. Chem. Phys. \textbf{147}, 234506 (2017).
\bibitem{IkedaZamponi} H. Ikeda and F. Zamponi,
arXiv:1812.08780.
\bibitem{BritoLernerWyart} C. Brito, E. Lerner, and M. Wyart,
Phys. Rev. X \textbf{8}, 031050 (2018).
\bibitem{Kapteijns} 
G. Kapteijns, W. Ji, C. Brito, M. Wyart, and E. Lerner,
Phys. Rev. E \textbf{99} 012106 (2019).
\bibitem{MDswap}
L. Berthier, E. Flenner, C.J. Fullerton, C. Scalliet, and M. Singh,
arXiv:1811.12837.
\bibitem{Szamelswap} G. Szamel,
Phys. Rev. E \textbf{98}, 050601(R) (2018).
\bibitem{Goetzebook}
W. G\"otze, \textit{Complex dynamics of glass-forming
liquids: A mode-coupling theory} (Oxford University Press, Oxford, 2008).
\bibitem{HansenMcDonald} J.P. Hansen and I.R. McDonald,
\textit{Theory of Simple Liquids}
(Elsevier, Amsterdam, 2006).
\bibitem{CHess} B. Cichocki and W. Hess, Physica A \textbf{141}, 475 (1987).
\bibitem{Kawasaki}K. Kawasaki, Physica A \textbf{215}, 61 (1995).
\bibitem{SL} G. Szamel and H. L\"{o}wen, Phys. Rev. A \textbf{44}, 8215 (1991).
\bibitem{PY1} J.L. Lebowitz and J. S. Rowlinson, J. Chem. Phys. \textbf{41}, 133
(1964).
\bibitem{PY2} R. J. Baxter, J. Chem. Phys. \textbf{52}, 4559 (1970).
\bibitem{Voigtmannthesis} We used explicit formulas from
Th. Voigtmann, Ph.D. thesis, TU M\"unchen, 2002.
\bibitem{GoetzeVoigtmann} W. G\"otze and Th. Voigtmann,
Phys. Rev. E \textbf{67}, 021502 (2003).
\bibitem{details} We discretized the self-consistent equations (\ref{arrest}) using a 
grid of 401 equally spaced wavevectors, with $q_\text{min}d_S=0$ and 
$q_\text{max}d_S=80$, where $d_S$ is the diameter of the smaller sphere. 
The resulting equations were then solved by iteration, with integrations performed 
using a gaussian quadrature.
\end{thebibliography}
\end{document}